\documentclass[aps,pra,showkeys,twocolumn,superscriptaddress]{revtex4-2}
\usepackage{array}
\usepackage{booktabs}
\usepackage{tabu}
\usepackage{dcolumn}
\usepackage{floatrow}
\usepackage{amsmath}
\usepackage{amsfonts}
\usepackage{float}
\usepackage{amssymb}
\usepackage{graphicx,color}
\usepackage[colorlinks={true}]{hyperref}
\hypersetup{colorlinks=true,linkcolor=red,citecolor=blue,urlcolor=blue}
\usepackage{graphicx}
\usepackage{dcolumn}
\usepackage{bm}
\usepackage{pstricks}
\usepackage{braket}
\usepackage{orcidlink}

\def\be{\begin{equation}}
  \def\ee{\end{equation}}
\def\bea{\begin{eqnarray}}
\def\eea{\end{eqnarray}}
\def\f{\frac}
\def\n{\nonumber}
\def\l{\label}
\def\p{\phi}
\def\o{\over}
\def\R{\rho}
\def\pa{\partial}
\def\om{\omega}
\def\na{\nabla}
\def\P{\Phi}
\begin{document}

\title{Steady-state work extraction from  two coupled qubits embedded within equilibrium
and non-equilibrium reservoirs} 

\author{Maryam Hadipour \orcidlink{0000-0002-6573-9960}}
\affiliation{Faculty of Physics, Urmia University of Technology, Urmia, Iran}

\author{Soroush Haseli \orcidlink{0000-0003-1031-4815}}\email{soroush.haseli@uut.ac.ir}
\affiliation{Faculty of Physics, Urmia University of Technology, Urmia, Iran}

\date{\today}
\def\be{\begin{equation}}
  \def\ee{\end{equation}}
\def\bea{\begin{eqnarray}}
\def\eea{\end{eqnarray}}
\def\f{\frac}
\def\n{\nonumber}
\def\l{\label}
\def\p{\phi}
\def\o{\over}
\def\R{\rho}
\def\pa{\partial}
\def\om{\omega}
\def\na{\nabla}
\def\P{$\Phi$}

\begin{abstract}
Work extraction is a fundamental aspect in thermodynamics. In the context of quantum physics, ergotropy quantifies the maximum amount of work that can be obtained from quantum system through cyclic unitary process. In this work, the steady-state ergotropy of two coupled qubit, each interacting locally with its individual boson or fermion reservoir, will be examined. In this work, both equilibrium and non-equilibrium scenarios for bosonic and fermionic environments interacting with the qubits will be considered.  In scenarios where two coupled qubits embedded within equilibrium boson  reservoirs, it has been observed that the  temperature of the reservoirs and the inter-qubits interaction strangth  act as detrimental factors in work extraction. In the case of fermionic equilibrium reservoirs, it will be observed that ergotropy grows monotonically with the reservoirs chemical potential. In the non-equilibrium boson reservoirs, the temperature difference between the two reservoirs is a destructive factor for ergotropy. In non-equilibrium fermion reservoirs, the situation is somewhat more complicated. For r base chemical potential values that are smaller than the qubit transition frequency, the behavior of ergotropy is non-monotonic. However, for base chemical potential values that are larger than the transition frequency, ergotropy grows monotonically with the reservoirs chemical potential difference. Also, we study the situation in which the coupled qubits are asymmetric. It is observed that the maximum work will be extracted in the situation where the coupled qubits  within both boson and fermion reservoirs be symmetric . 
\end{abstract}
\keywords{Equilibrium, Non-equilibrium, Ergotropy, Coupled qubits }

\maketitle

\section{Introduction}\label{intro}
Historically, thermodynamics has been associated with systems composed of a vast number of particles. Thermodynamics is a fundamental component of our current understanding of the physical world. It has stayed the same even through major revolutions in physics, such as relativity and quantum theory. Thermodynamics was initially introduced to study thermal systems on a macroscopic scale, before the advent of quantum mechanics at the microscopic level. However, with the advent of quantum mechanics, clear differences are observed in dealing with microscopic thermal systems compared to macroscopic systems. In the quantum regime, the behavior of systems is governed by quantum mechanics, where the principles of superposition, entanglement, and quantization come into play \cite{1,2,3,4,5,6,7}. The laws of traditional thermodynamics, when applied within the realm of quantum mechanics, have led to the emergence of a new field in quantum theory known as quantum thermodynamics \cite{8,9,10,11,12,13}. In quantum thermodynamics ,efficient work extraction from quantum systems is a fundamental goal in studying quantum biological systems and modern nanoscale technologies.  To determine the amount of extractable work  from a quantum system, the concept of "ergotropy" is commonly used. Ergotropy refers to the maximum amount of work that can be extracted from a quantum system trough employing the cyclic unitary operations \cite{14}.

In practical scenarios, a quantum system unavoidably engages with its surrounding environments. So, study the open quantum systems has much attraction in quantum information theory \cite{15}. The convergence of the concepts of open quantum systems and quantum thermodynamics has led to the expansion of operational quantum devices, such as  quantum heat engines \cite{16,17,18}, quantum refrigerators \cite{19} and quantum batteries \cite{20,21,22,23,24,25,26,27,28,29,30,31,32,33,34,35,36,37,38,39,40,41,42,43,44,45,46,47,48,49,50,51,52,53,54,55,56,57,58,59,60,61,62,63,64,65,66,67}. 
It is thus critical to study the work extraction in the context of open quantum systems. In recent years, there has been an increasing fascination in exploring open quantum systems within non-equilibrium settings \cite{68,69,70,71}. The environments remain out of equilibrium due to a steady temperature difference or chemical potential difference. This causes energy or matter flow through the quantum system and environments, continuously keeping them away from thermodynamic equilibrium \cite{72,73,74,75}. The simplest model imaginable for studying  extractable work in non-equilibrium steady states is likely a two-qubit system coupled with two reservoirs \cite{76,77}. In this work, the steady-state work extraction of the two-qubit system interacting with two independent boson or fermion reservoirs that can exchange energy (boson reservoir) or particles (fermion reservoir) with the system in both equilibrium and non-equilibrium settings is investigated \cite{78}. In our study, the bosonic and fermionic environments are considered for both equilibrium and non-equilibrium settings. In this work, the extractable work from the two asymmetric qubits system for both equilibrium and non-equilibrium bosonic and fermionic reservoirs will also be examined. 

The work is organized as follows. In Sec.\ref{Model}, the model and methods that uses in this work will be introduced. In Sec.\ref{ergot}, ergotropy in the equilibrium reservoirs will be studied. Ergotropy in the equilibrium reservoirs will be examined in Sec.\ref{ergotn}. In Sec. \ref{sym}, work extraction from asymmetric qubits will be investigated. In Sec. \ref{conclusion}, the results of the study are summarized.  
\section{Models and methods}\label{Model}
\begin{figure}[!h]
    \centering
  \includegraphics[width = 0.85\linewidth]{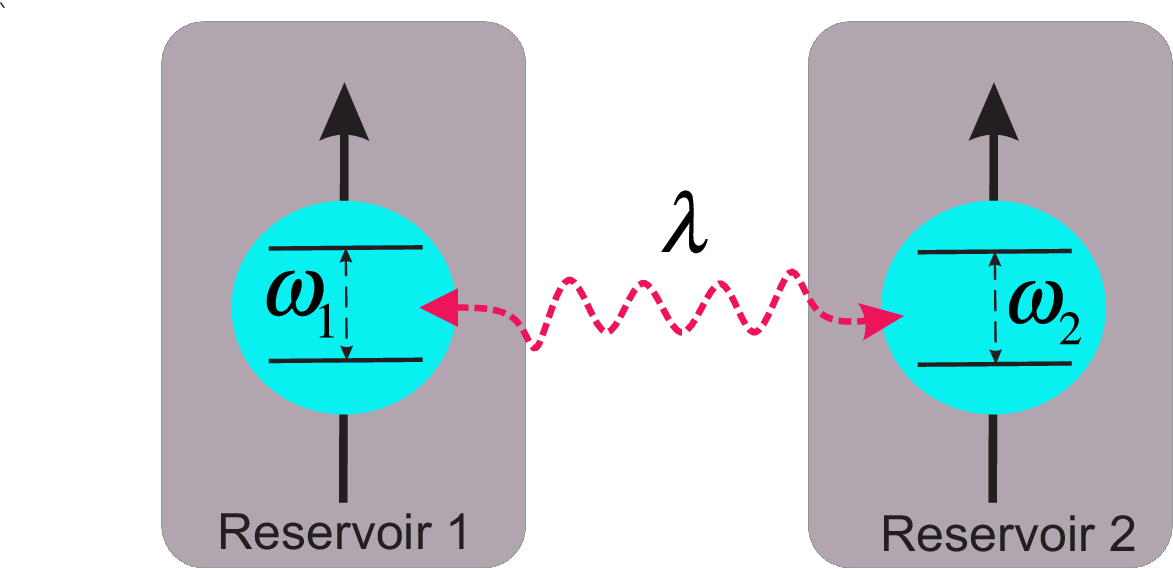}
    \centering
    \caption{The schematic diagram of the considered model includes two interacting qubits in two separate reservoirs. $\lambda$ represents the inter-qubit coupling strength, and $\omega_1$ and $\omega_2$ are the energy level differences between the ground and excited states for the qubits.}\label{Fig1}
\end{figure}

The schematic diagram of the model is represented in Fig. \ref{Fig1}. The considered model consists of two qubits that are coupled together, with each qubit placed in its own environment. The environments follow either fermionic or bosonic statistics. The total Hamiltonian of the system includes the coupled qubits and environment can be written as $H=H_S+H_R+H_{int}$ where $H_S$, $H_R$ and $H_{int}$ are the system, reservoir and interaction terms of the total Hamiltonian respectively. They can be written as 
\begin{equation}\label{Hamiltonian1}
\begin{gathered}
H_s=\omega_1|e\rangle_1\left\langle e\left|+\omega_2\right| e\right\rangle_2\langle e|+\frac{\lambda}{2}\left[\sigma_{+}^{(1)} \sigma_{-}^{(2)}+\sigma_{-}^{(1)} \sigma_{+}^{(2)}\right], \\
H_R=\sum_k \omega_{1 k} b_k^{\dagger} b_k+\sum_k \omega_{2 k} c_k^{\dagger} c_k, \\
V=\sum_k g_{1k}\left[\sigma_{-}^{(1)} b_k^{\dagger}+\sigma_{+}^{(1)} b_k\right]+\sum_k g_{2k}\left[\sigma_{-}^{(2)} c_k^{\dagger}+\sigma_{+}^{(2)} c_k\right]
\end{gathered}
\end{equation}

\begin{figure}[!h]
    \centering
  \includegraphics[width = 0.85\linewidth]{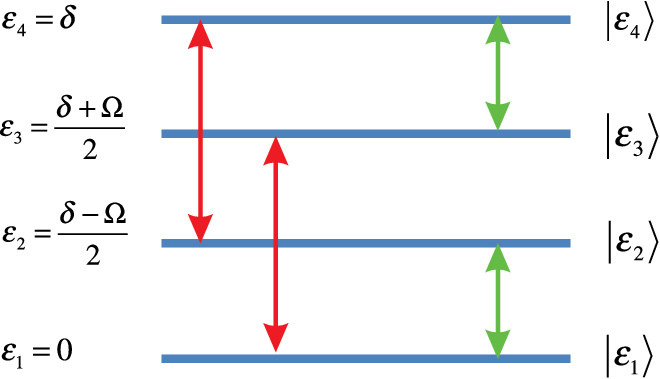}
    \centering
    \caption{The schematic diagram of eigenenergies $\varepsilon_i$ and the corresponding eigenstates $\vert \varepsilon_i \rangle$ of the
Hamiltonian for the coupled qubit system with $i=1,2,3,4$.}\label{Fig2}
\end{figure}

where  $\omega_1$ and $\omega_2$ are the transition frequencies of the qubits, $\lambda$ denotes the coupling strength between qubits. $b_k$ ($b_k^\dag$) and $c_k$($c_k^\dag$) are the annihilation (creation) operators for the $k$th mode of the reservoir with frequencies $\omega_{1k}$ and $\omega_{2k}$ which are respectively in contact with qubits $1$ and $2$.  $g_{1k}$ and $g_{2k}$ denote the coupling strength between reservoir and qubits $1$ and $2$ respectively. The eigenvalues and the corresponding eigenvectors of the coupled qubit Hamiltonian are obtained as follows 
\begin{eqnarray}
\varepsilon_1 &=& 0, \qquad \vert \varepsilon_1 \rangle = \vert gg \rangle, \\
\varepsilon_2 &=& \frac{\delta -\Omega}{2}, \qquad \vert \varepsilon_2 \rangle = -\sin \frac{\theta}{2} \vert eg \rangle + \cos \frac{\theta}{2} \vert ge \rangle,  \nonumber \\
\varepsilon_3 &=& \frac{\delta + \Omega}{2}, \qquad \vert \varepsilon_2 \rangle = \cos \frac{\theta}{2} \vert eg \rangle -\sin \frac{\theta}{2} \vert ge \rangle,   \nonumber \\
\varepsilon_4 &=&\delta, \qquad \vert \varepsilon_4 \rangle = \vert ee \rangle, \nonumber 
\end{eqnarray}
where $\delta=\omega_1 + \omega_2$, $\Omega=\sqrt{\Delta^2 + \lambda^2}$ is the Rabi frequency  with $\Delta=\omega_1 - \omega_2$ and $\theta \in [0, \pi ]$ is the mixing angle which is defined by $\theta = \arctan(\lambda /\Delta)$. In the case of symmetric qubits, i.e., when $\Delta=0$, the mixing angle  is $\theta=\pi/2$. In the case of asymmetric qubits  i.e. $\omega_1 \neq \omega_2$, the mixing angle is equal to $\theta = \arctan(\lambda /\Delta)$ when $\omega_1 > \omega_2$ and $\theta = \pi+ \arctan(\lambda /\Delta)$ when $\omega_1 < \omega_2$. In order to ensure the rotating wave approximation in the qubit reservoir interaction Hamiltonian, it is necessary to have $\lambda < 2 \sqrt{\omega_1 \omega_2}$. This requirement ensures that $\delta > \Omega $ and allows the eigenenergies to be ordered as $\varepsilon_1 < \varepsilon_2 < \varepsilon_3 < \varepsilon_4$. Fig.\ref{Fig2} shows the schematic diagram of eigenenergies and the corresponding eigenstates of the
Hamiltonian for the coupled qubit system.
\vspace{0.2 cm}

The unitary transformation between the energy eigenbasis $\left\lbrace \vert \varepsilon_1 \rangle,  \vert \varepsilon_2 \rangle,  \vert \varepsilon_3 \rangle,  \vert \varepsilon_4 \rangle \right\rbrace $ and the standard eigenbasis $ \left\lbrace  \vert ee \rangle, \vert eg \rangle, \vert ge \rangle, \vert gg \rangle \right\rbrace $ is defined as 
\begin{equation}\label{unit}
U=\left(\begin{array}{cccc}
1 & 0 & 0 & 0  \\
0 & \cos  \frac{\theta}{2} & \sin \frac{\theta}{2} & 0 \\
0 & -\sin \frac{\theta}{2} & \cos \frac{\theta}{2} & 0  \\
0 & 0 & 0 & 1
\end{array}\right).
\end{equation}
The interaction term of the total Hamiltonian in energy eigenbasis can be rewritten as  
\begin{equation}
H_{int}= \sum_k [g_{1k} (A_1 +B_1) b_k^\dag + g_{2k}(A_2 + B_2)c_k^\dag] + H.c.,
\end{equation}
where 
\begin{eqnarray}
A_1 &=& \sin \frac{\theta}{2}(\vert \varepsilon_3 \rangle \langle \varepsilon_4 \vert - \vert \varepsilon_1 \rangle \langle \varepsilon_2 \vert ), \\
B_1 &=& \cos   \frac{\theta}{2} ( \vert \varepsilon_2 \rangle \langle \varepsilon_4 \vert + \vert \varepsilon_1 \rangle \langle \varepsilon_3 \vert), \nonumber \\
A_2 &=& \cos \frac{\theta}{2}(\vert \varepsilon_3 \rangle \langle \varepsilon_4 \vert + \vert \varepsilon_1 \rangle \langle \varepsilon_2 \vert ), \nonumber \\
B_2 &=& \sin   \frac{\theta}{2} (  \vert \varepsilon_1 \rangle \langle \varepsilon_3 \vert - \vert \varepsilon_2 \rangle \langle \varepsilon_4 \vert). \nonumber
\end{eqnarray}
Considering free Hamiltonian $H_0=H_S+H_R$ and using interaction picture we have 
\begin{eqnarray}\label{Hint}
H_{int}(t)&=& \sum_k g_{1k} \left( A_1 e^{-i \frac{\delta-\Omega}{2}t}+ B_1 e^{-i \frac{\delta+\Omega}{2}t } \right) b_k^\dag e^{i \omega_{1k}t}     \\
  &+& \sum_k g_{2k} \left( A_2 e^{-i \frac{\delta-\Omega}{2}t}+ B_2 e^{-i \frac{\delta+\Omega}{2}t } \right) c_k^\dag e^{i \omega_{2k}t} + H.c. \nonumber
\end{eqnarray}
In the interaction picture, the quantum master equation under the Born-Markov approximation is given by \cite{15}
\begin{equation}
\frac{d \rho_I(t)}{dt}= - \int_0^{\infty} ds Tr_B \left[ H_{int}(t),\left[H_{int}(t-s), \rho_I(t)\otimes (t) \rho_B \right]  \right], 
\end{equation}
where $\rho_I(t)$ and $\rho_B$ are the reduced density operation of the two-qubit system in interaction picture and density operator of the reservoirs at equilibrium state. In this step, returning to the Schrödinger picture and avoiding the secular approximation, ultimately one can obtain the master equation which is referred to as the Bloch-Redfield equation for the system's density matrix as 
\begin{equation}\label{masterw}
\frac{d \rho}{dt} = -i [H_S,\rho]+ \mathcal{D}_1 [\rho] + \mathcal{D}_2[\rho],
\end{equation}
where $\mathcal{D}_1[\rho]=\sum_{j=1}^{2} \mathcal{N}_j(\rho)$ and  $\mathcal{D}_2[\rho]=\sum_{j=1}^2 \mathcal{S}_i(\rho)$ with 
\begin{equation}
\begin{aligned}
\mathcal{N}_i[\rho]= & \gamma_i^{+}\left[2 B_i^{\dagger} \rho B_i-B_i B_i^{\dagger} \rho-\rho B_i B_i^{\dagger}\right] \\
& +\gamma_i^{-}\left[2 A_i^{\dagger} \rho A_i-A_i A_i^{\dagger} \rho-\rho A_i A_i^{\dagger}\right] \\
& +\Gamma_i^{+}\left[2 B_i \rho B_i^{\dagger}-B_i^{\dagger} B_i \rho-\rho B_i^{\dagger} B_i\right] \\
& +\Gamma_i^{-}\left[2 A_i \rho A_i^{\dagger}-A_i^{\dagger} A_i \rho-\rho A_i^{\dagger} A_i\right],
\end{aligned}
\end{equation}
and 
\begin{equation}
\begin{aligned}
\mathcal{S}_i[\rho]= & \gamma_i^{+}\left[A_i^{\dagger} \rho B_i+B_i^{\dagger} \rho A_i-A_i B_i^{\dagger} \rho-\rho B_i A_i^{\dagger}\right] \\
& +\gamma_i^{-}\left[A_i^{\dagger} \rho B_i+B_i^{\dagger} \rho A_i-B_i A_i^{\dagger} \rho-\rho A_i B_i^{\dagger}\right] \\
& +\Gamma_i^{+}\left[A_i \rho B_i^{\dagger}+B_i \rho A_i^{\dagger}-A_i^{\dagger} B_i \rho-\rho B_i^{\dagger} A_i\right] \\
& +\Gamma_i^{-}\left[A_i \rho B_i^{\dagger}+B_i \rho A_i^{\dagger}-B_i^{\dagger} A_i \rho-\rho A_i^{\dagger} B_i\right]
\end{aligned}
\end{equation}
where $\gamma^{\pm}\equiv \gamma_i (\delta /2 \pm \Omega /2)$ and $\Gamma^{\pm} \equiv \Gamma_i (\delta /2 \pm \Omega/2)$. In bosonic reservoir we have 
\begin{equation}\label{bos}
\gamma_i (\omega) = J_i (\omega) N_i
(\omega), \quad \Gamma_i(\omega)[N_i(\omega)+1],
\end{equation}
while for fermionic reservoirs we have 
\begin{equation}\label{fer}
\gamma_i (\omega) = J_i (\omega) N_i
(\omega), \quad \Gamma_i(\omega)[1-N_i(\omega)].
\end{equation}
In above equations, $J_i(\omega)=\pi \sum_k g_{ik}^2 \delta(\omega - \omega_{ik})$ is the spectral density of the $i$th reservoir \cite{78}. The average particle number at a given frequency $\omega$ in the $i$th reservoir can be  described by $N_i(\omega)= (e^{\frac{\omega-\mu_i}{T_i}}\pm 1)^{-1}$, where minus sign is used for boson reservoir with Bose-Einstein statistics and plus sign is used for fermion reservoir with Fermi-Dirac statistics. Here, $\mu_i$ denotes the chemical potential and $T_i$ represents the temperature of the $i$th reservoir. In practical scenarios involving boson reservoirs, such as photon or phonon baths, the number of particles is generally not conserved. So, the chemical potential becomes zero for boson reservoir. This leads to the expression $N_i(\omega)= (e^{\frac{\omega}{T_i}}\pm 1)^{-1}$ for boson reservoir. For fermionic reservoirs, the chemical potential is maintained , meaning the system can exchange particles with the fermionic environment during the process i.e we have $\mu_i \neq 0$. At this stage, we are dealing only with spectral densities that are balanced and frequency-independent $J_1(\delta/2 \pm \Omega/2)=J_2(\delta/2 \pm \Omega/2)=J$. To gain a clearer understanding about physical processes outlined by the quantum master equation, we observe that the structure of the interaction Hamiltonian $H_{int}$ in Eq.\ref{Hint} indicates that the interaction between the two qubit system and the reservoirs causes two distinct sets of energy-level transitions within the system, as illustrated in Fig.\ref{Fig2}. One set includes transitions $\vert \varepsilon_1 \rangle \leftrightarrow \vert \varepsilon_2 \rangle$ and $\vert \varepsilon_3 \rangle \leftrightarrow \vert \varepsilon_4 \rangle$ with transition frequency $(\delta - \Omega)/2$, which are represented by green arrow and the other set includes transitions  $\vert \varepsilon_2 \rangle \leftrightarrow \vert \varepsilon_4 \rangle$  and $\vert \varepsilon_1 \rangle \leftrightarrow \vert \varepsilon_3 \rangle$ with transition frequency $(\delta + \Omega)/2$, which are showed by red arrow.  The dissipator term $\mathcal{D}_0[\rho]$ characterizes processes where the energy emitted to the environment by the system, due to an energy-level transition, is reabsorbed by the transitions within the same group at the same frequency. The dissipator term $\mathcal{D}_b[\rho]$ shows the processes where the emission and reabsorption of energy are enabled by energy-level transitions in different groups with diferent process frequencies. The process associated with $\mathcal{D}_b[\rho]$ is typically regarded as a rapidly oscillating process and is eliminated using the secular approximation. The secular approximation works well at equilibrium situations. ($T_1=T_2$) and ($\mu_1=\mu_2, T_1=T_2$) are the equilibrium situations for boson and fermion reservoirs respectively. Under these circumstances, the diagonal components of the density matrix become independent of the off-diagonal ones. The density matrix in the equilibrium steady state is diagonal, devoid of any remaining coherence in the energy eigenstate depiction. However, the non-equilibrium situation for bosonic reservoirs is $T_1  \neq T_2$ and for fermion reservoir is $\mu_1 \neq \mu_2$ and $T_1 \neq T_2$.

The steady state density matrix can be obtain by solving $d \rho /dt = 0$.  Here, we just present the general form of the steady-state density matrix that results from the solution of $d \rho /dt = 0$. For more details on solving this equation, refer to Ref. \cite{78}. In the absence of the secular approximation, the diagonal elements of the steady-state density matrix are coupled with the diagonal elements of the density matrix. Hence, the steady state density matrix in energy eigenstate $\rho_{\varepsilon}$ can be obtained as 
\begin{equation}
\rho_{\varepsilon}=\left(\begin{array}{cccc}
\rho_{11} & 0 & 0 & 0  \\
0 & \rho_{22} & \rho_{23} & 0 \\
0 & \rho_{32} & \rho_{33} & 0  \\
0 & 0 & 0 & \rho_{44}
\end{array}\right).
\end{equation}
Using Eq.\ref{unit}, the steady state density matrix in general basis $ \left\lbrace  \vert ee \rangle, \vert eg \rangle, \vert ge \rangle, \vert gg \rangle \right\rbrace $ can be obtained as 
\begin{equation}
\rho=\left(\begin{array}{cccc}
\eta_{11} & 0 & 0 & 0  \\
0 & \eta_{22} & \eta_{23} & 0 \\
0 & \eta_{32} & \eta_{33} & 0  \\
0 & 0 & 0 & \eta_{44}
\end{array}\right).
\end{equation}
where the element of the density matrix $\rho$ are 
\begin{eqnarray}\label{dm}
\eta_{11}&=&\rho _{11}, \quad \eta_{44}=\rho _{44},\\ 
\eta_{22}&=&\frac{1}{2} \left(\left(\rho _{23}+\rho _{32}\right) \sin \theta +\rho _{22} (\cos \theta +1)-\rho _{33} (\cos \theta -1)\right), \nonumber \\
\eta_{33}&=&\frac{1}{2} \left(\rho _{33} (\cos \theta +1) -\left(\rho _{23}+\rho _{32}\right) \sin \theta -\rho _{22} (\cos  \theta +1) \right), \nonumber\\
\eta_{23}&=&\frac{1}{2} \left(\left(\rho _{33}-\rho _{22}\right) \sin \theta +\rho _{23} (\cos  \theta +1)+\rho _{32} (\cos  \theta -1)\right), \nonumber
\end{eqnarray}
Here, we aim to examine the maximum extractable work from the steady-state system under different environmental situations. Therefore, before anything else, we will briefly review the concept of ergotropy. Ergotropy is defined as the highest amount of energy that can be obtained from a quantum system via a cyclic unitary operation with unitary operation $U$ \cite{14}. So, the ergotropy can be given by $\mathcal{E}(\rho)= \mathcal{U}(\rho)-\min_U Tr(H_s U \rho U^{\dag})$, where $\mathcal{U}(\rho)=Tr(\rho H_s)$ is the internal energy of the system  and $H$ is the Hamiltonian of the considered system. In mentioned ergotropy formula the optimization is taken over the all set of the unitary operations $U$. It has been demonstrated that for any arbitrary state $\rho$, there is a unique state $\mathcal{P}_\rho$ that maximizes the given equation. The state $\mathcal{P}_\rho$ is called passive state. Hence, the ergotropy can be rewritten as $\mathcal{E}(\rho)=\mathcal{U}(\rho)-Tr(P_\rho H_s)$. Now, let's consider the spectral decomposition of the density matrix $\rho$  and the system Hamiltonian $H_s$  as follows
\begin{eqnarray}
\rho &=& \sum_n r_n \vert r_n \rangle \langle r_n \vert, \quad r_n \geq r_{n+1}, \nonumber \\
H_s &=& \sum_m \varepsilon_m \vert \varepsilon_m \rangle \langle \varepsilon_m \vert,  \quad \varepsilon_m \leq \varepsilon_{m+1}.
\end{eqnarray}
From above the passive state can be given by $\mathcal{P}_\rho=\sum_n r_n \vert \varepsilon_n \rangle \langle \varepsilon_n \vert$, where $r_n$ ($\vert r_n \rangle$) and  $\varepsilon_m$ ($\vert \varepsilon_m \rangle$) are eigenvalues (eigenstates) of the density matrix $\rho$ and Hamiltonian $H_s$ respectively. So, the close form for ergotropy $\mathcal{E}(\rho)$ can be obtained as
\begin{equation}\label{tergo}
\mathcal{E}(\rho)= \sum_{n,m} r_n \varepsilon_m \left( \left| \langle r_n \vert \varepsilon_m \rangle \right|^2 - \delta_{m,n}  \right), 
\end{equation}
where $\delta_{m,n}$ is the Kronecker delta function.
\section{Ergotropy  IN THE EQUILIBRIUM SITUATION}\label{ergot}
In equilibrium, both reservoirs have the same temperature and chemical potential i.e. $T_1=T_2$ and $\mu_1=\mu_2$. Here, the symmetric case is considered for the coupled qubits pair. So, we have $\omega_1=\omega_2=\omega$ and the mixing angle $\theta$ is equal to $\pi/2$. In this section, we aim to enhance the physical understanding of work extraction in equilibrium settings, areas that have been less explored in previous researches. At the the equilibrium steady state the off-diagonal element of density matrix will be vanished. The diagonal element can be obtained as \cite{79}
\begin{eqnarray}\label{element}
\rho_{11}&=& \frac{(\Gamma_1^+ + \Gamma_2^+)(\gamma_1^- + \gamma_2^-)}{\mathcal{Z}},        \\
\rho_{22}&=&    \frac{(\gamma_1^- + \gamma_2^-)(\gamma_1^+ + \gamma_2^+)}{\mathcal{Z}},          \nonumber \\
\rho_{33}&=&           \frac{(\Gamma_1^- + \Gamma_2^-)(\gamma_1^+ + \gamma_2^+)}{\mathcal{Z}}, \nonumber \\
\rho_{44}&=&    \frac{(\Gamma_1^- + \Gamma_2^-)(\Gamma_1^+ + \Gamma_2^+)}{\mathcal{Z}},        \nonumber
\end{eqnarray}
where $\mathcal{Z}$ is normalization factor which can be written as
\begin{equation}
\mathcal{Z}=(\gamma_1^- + \gamma_2^- + \Gamma_1^- + \Gamma_2^-)(\gamma_1^+ + \gamma_2^{+}+\Gamma_1^+ + \Gamma_2^+).
\end{equation}
Now, we can proceed to study ergotropy for both bosonic and fermionic equilibrium reservoirs. 
\subsection{Ergotropy in equilibrium boson reservoirs}
In this section, we consider a scenario where the coupled qubits are influenced by bosonic equilibrium reservoirs with identical temperatures $T_1=T_2$.  Using Eq. \ref{bos}, it can be find that 
\begin{eqnarray}
\gamma_1^+ &=& \gamma_2^+ = \frac{J}{e^{\omega_+/T}-1}, \Gamma_1^+ = \Gamma_2^+=\frac{J e^{\omega_+/T} }{e^{\omega_+/T}-1}, \\
\gamma_1^- &=& \gamma_2^- = \frac{J}{e^{\omega_-/T}-1}, \Gamma_1^- = \Gamma_2^-=\frac{J e^{\omega_-/T} }{e^{\omega_-/T}-1}, \nonumber
\end{eqnarray}

\begin{figure}[!h]
    \centering
  \includegraphics[width = 0.85\linewidth]{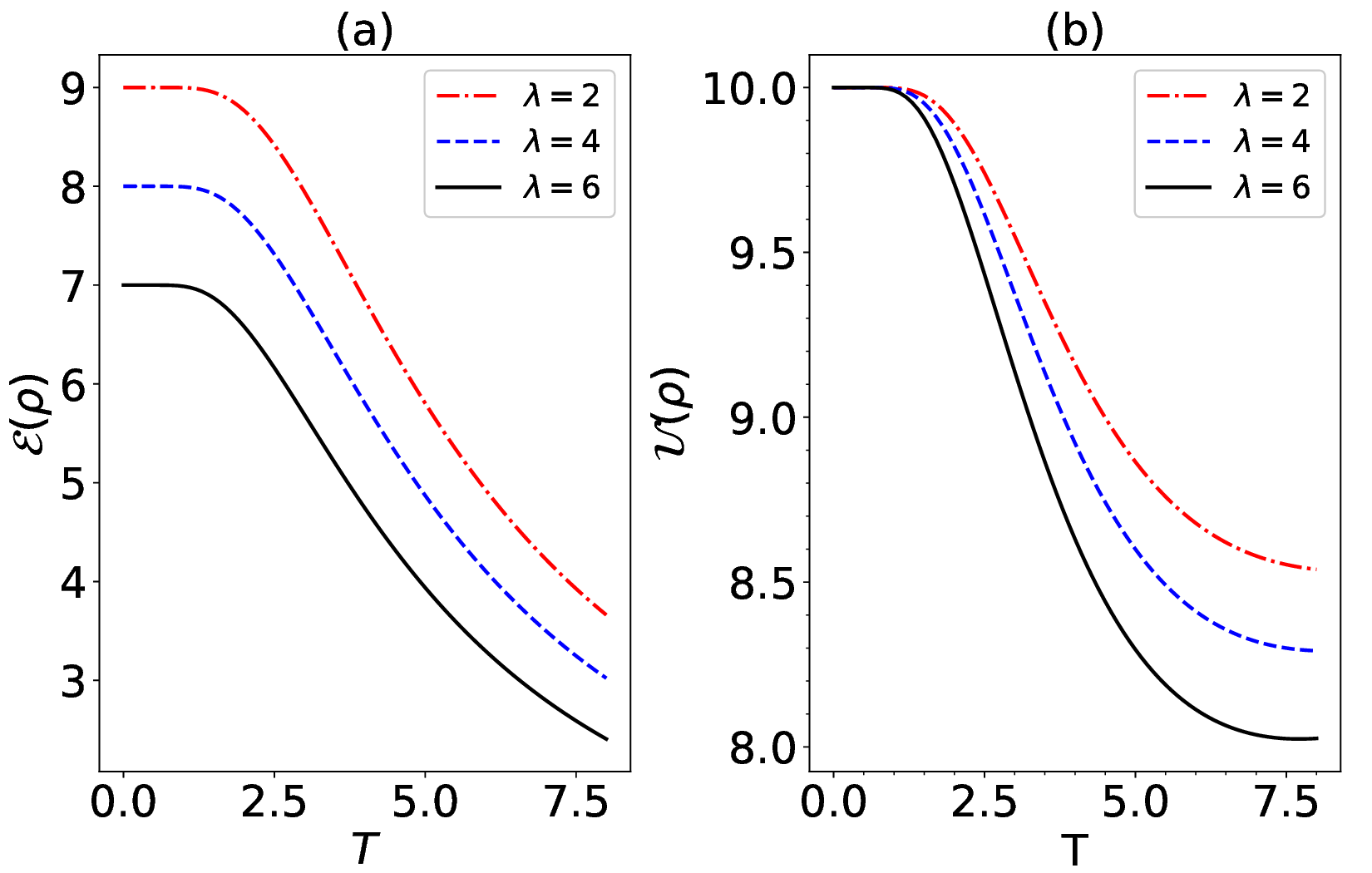}
    \centering
    \caption{(a) The steady-state ergotropy in equilibrium boson reservoir as a function of temperature $T$ for different value of the interqubit coupling strength $\lambda$. (b) The internal energy in equilibrium boson reservoir as a function of temperature $T$ for different value of the interqubit coupling strength $\lambda$. The values of other parameters for both plot are $\omega_1=\omega_2=10$, $J_1=J_2=1$.}\label{Fig3}
\end{figure}

where $\omega_\pm = \omega \pm \lambda/2$. So, using Eq.\ref{element}, the density matrix of steady state at equilibrium bosonic reservoir can be obtained. In Fig.\ref{Fig3}, steady-state ergotropy and internal energy in equilibrium boson reservoir have been plotted as a function of the temperature. It can be seen that the steady-state ergotropy and internal energy decreases with increasing the temperature of the bosonic reservoir. In Fig. \ref{Fig3}, the effect of the interqubit coupling strength on both steady-state ergotropy and internal energy has also been illustrated. It can be seen that as interqubit coupling strength increases, both staedy-state ergotropy and internal energy decrease. Our results  reveal that the thermal effect due to the reservoir temperatures reduces the maximum extractable work from the coupled qubits within an equilibrium boson reservoir. 
\subsection{Ergotropy in equilibrium fermion reservoirs} 
Due to the detrimental effects of temperature in the equilibrium boson  reservoirs on the maximum extractable work from the coupled qubits, in this part, we intend to examine the effect of particle exchange (chemical potential) between the reservoir and the pair of qubits placed in the equilibrium boson  reservoirs with $T_1=T_2=T$ and $\mu_1=\mu_2=\mu$. In accordance with what we had in the equilibrium boson  reservoirs, using Eq.\ref{fer} we will have 
\begin{eqnarray}
\gamma_1^+&=& \gamma_2^+ = \frac{J}{e^{\omega_{f+}/T}+1}, \Gamma_1^+=\Gamma_2^+ = \frac{J e^{\omega_{f+}/T}}{e^{\omega_{f+}/T}+1}, \\
\gamma_1^-&=& \gamma_2^- = \frac{J}{e^{\omega_{f-}/T}+1}, \Gamma_1^-=\Gamma_2^- = \frac{J e^{\omega_{f-}/T}}{e^{\omega_{f-}/T}+1}, \nonumber
\end{eqnarray}
where $\omega_{f \pm}=\omega - \mu \pm \lambda/2$.

\begin{figure}[!h]
    \centering
  \includegraphics[width = 0.85\linewidth]{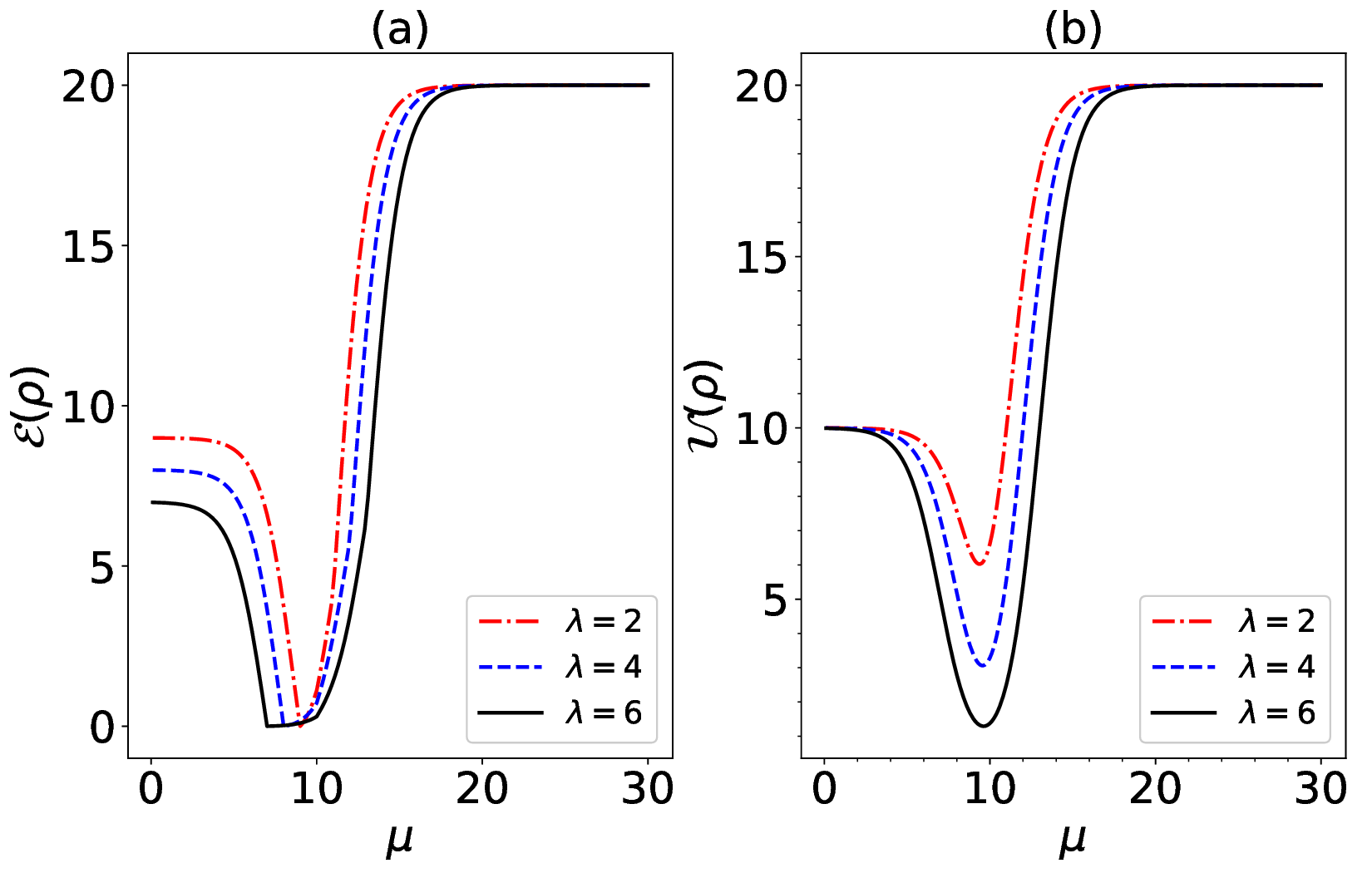}
    \centering
    \caption{(a) The steady-state ergotropy in equilibrium fermion reservoir as a function of chemical potential $\mu$ for different value of the interqubit coupling strength $\lambda$. (b) The internal energy in equilibrium fermion reservoir as a function of chemical potential $\mu$ for different value of the interqubit coupling strength $\lambda$. The values of other parameters for both plot are $\omega_1=\omega_2=10$, $J_1=J_2=1$ and $T=1.5$.}\label{Fig4}
\end{figure}

Fig.\ref{Fig4}, shows the steady state ergotropy and internal energy in terms of the chemical potential $\mu$ for different values of the inter-qubit interaction strength $\lambda$.  From Fig. \ref{Fig4}(a), It can be seen that the ergotropy  has a more complicated behavior with respect to chemical potential $\mu$. It can be seen that for small $\mu$, the ergotropy decreases with increasing the amount of the chemical potential $\mu$ while for large amount of the $\mu$ ($\mu > \omega$) the ergotropy increase with increasing the chemical potential $\mu$  and asymptotically reaches a maximum constant value. It is also observed that the ergotropy decreases with increasing the inter-qubit interaction strength. The same results can be obtained from Fig.\ref{Fig4}(b) for steady-state internal energy. 
\section{Ergotropy IN THE NONEQUILIBRIUM SITUATION}\label{ergotn}
In the previous section, we examined the steady-state ergotropy and internal energy under equilibrium conditions. In this section, we explore the steady state ergotropy and internal energy  at the nonequilibrium scenario where the two reservoirs have different temperatures and different chemical potentials forn boson and fermion reservoirs. Here, we proceed with the symmetric case for the coupled qubits with $\omega_1=\omega_2=\omega$ ($\theta=\pi/2$, $\delta= 2 \omega$, $\Omega=\lambda$ ). In Ref. \cite{78} , the nonequilibrium density matrix has been obtained analytically. Here, we briefly review it. The following notation is used to write the nonequilibrium density matrix : 
\begin{eqnarray}
L_+ &=& \frac{1}{2}(N_1^+ + N_2^+), \quad L_-=\frac{1}{2}(N_1^- + N_2^-), \\
M_+ &=&  \frac{1}{2}(N_1^+ - N_2^+), \quad M_-=\frac{1}{2}(N_1^- - N_2^-), \nonumber
\end{eqnarray}
where $N_i^{\pm}=N_i(\delta/2 \pm \Omega/2)$ with $N_i(\omega)$ is the average particle number at a given frequency $\omega$ in the $i$th reservoir. In the above formulation, $M_\pm$ indicates the non-equilibrium nature of the setting. This means that for an equilibrium case, $M_\pm=0$ and for a non-equilibrium case $M_\pm \neq 0$. However, $L_\pm$ which represents the average particle number between the two reservoirs, reflects the average equilibrium effect of the two reservoirs.
\subsection{Ergotropy in non-equilibrium boson reservoirs}
Let us consider the coupled qubits, each embedded individually in its  bosonic reservoir. In this nonequilibrium scenario, the two reservoirs have different temperatures $T_1 \neq T_2$, and there exists a temperature difference $\Delta T =T_2 -T_1$ between reservoirs. The elements of the steady-state density matrix in the energy eigenstate can be obtained as 
\vspace{-0.3cm}
\begin{eqnarray}
\rho_{11}&=& \frac{1}{\mathcal{N}}\left[ (1+L_+)(1+L_-)- s_1 s_2 R \right], \\ 
\rho_{22}&=&  \frac{1}{\mathcal{N}}\left[ L_-(1+L_+)+s_2 z_1 R\right], \nonumber \\
\rho_{33}&=&  \frac{1}{\mathcal{N}}\left[ L_+(1+L_-)+s_1 z_2 R\right], \nonumber \\
\rho_{44}&=&  \frac{1}{\mathcal{N}}\left[ L_+L_--z_1z_2 R \right], \nonumber \\
\rho_{23}&=& \frac{1}{\mathcal{N}} \left[  \frac{M_+(1+2L_-)+M_-(1+2L_+)}{2(1+L_++L_-)+i\frac{\Omega}{J}}\right], \nonumber
\end{eqnarray}
where 
\begin{eqnarray}
s_1 &=& M_+-M_-(3+2L_++2 L_-), \\ 
s_2 &=& M_--M_+(3+2L_++2 L_-), \nonumber \\
z_1 &=& M_+ + M_-(1+2 L_++2L_-), \nonumber \\
z_2 &=& M_- + M_+(1+2 L_++2L_-), \nonumber \\
R &=& \frac{1}{4(1+L_++L_-)^2 + (\frac{\Omega}{J})^2}.
\end{eqnarray}
In above relations $\mathcal{N}$ is the normalization factor that can be given by 
\begin{eqnarray}
\mathcal{N}&=&(1+2 L_+)(1+ 2L_-) \\
&-& 16 M_+ M_-(1+L_++L_-)^2 R. \nonumber 
\end{eqnarray}

\begin{figure}[!h]
    \centering
  \includegraphics[width = 0.85\linewidth]{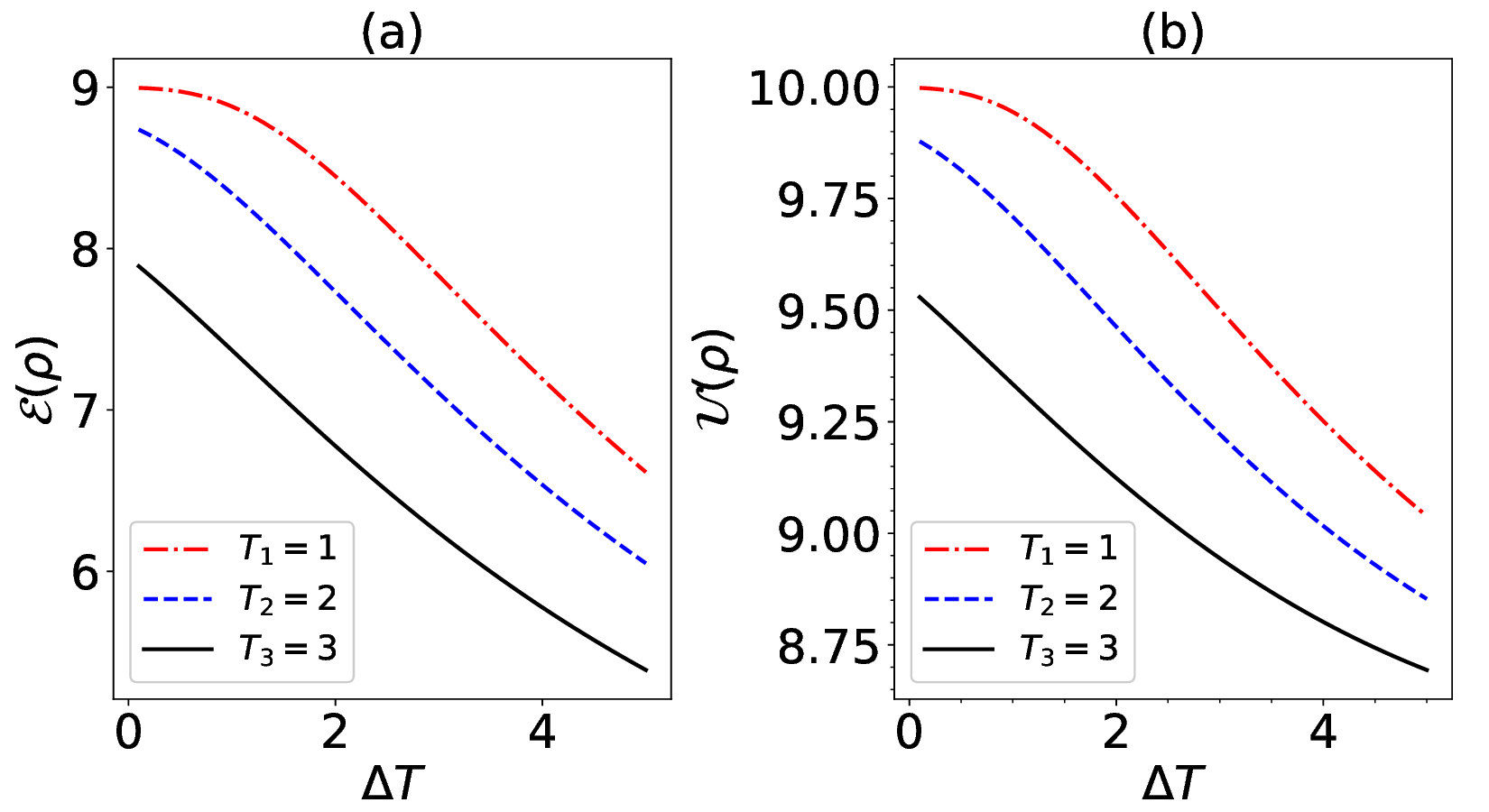}
    \centering
    \caption{(a) The steady-state ergotropy in non-equilibrium boson reservoir as a function of $\Delta T$ for different value of the base temperature $T_1$. (b) The internal energy in non-equilibrium boson reservoir as a function of  $\Delta T$ for different value of the base temperature $T_1$. The values of other parameters for both plot are $\omega_1=\omega_2=10$, $J_1=J_2=1$ and $\lambda=2$.}\label{Fig5}
\end{figure}

In Fig.\ref{Fig5}, variations of the steady state ergotropy and internal energy are depicted as a function of the temperature difference $\Delta T$ between the two boson reservoir for different value of the base temperature $T_1$. From Fig. \ref{Fig5}(a), it can be seen that the steady state ergotropy in non-equilibrium case decreases with increasing the difference temperature between individual reservoir. It can also be seen that the steady state ergotropy decreases with increasing the value of the base temperature $T_1$. The results shows that the temperature has destructive effect on the work extraction from the coupled qubits system in its individual reservoir with different temperature. The same results are observed for steady state internal energy from Fig.\ref{Fig5}(b). 
\subsection{Ergotropy in non-equilibrium fermion reservoirs}
Let us consider the coupled qubits, each embedded individually in its  bosonic reservoir with equal temperature $T_1=T_2$ and different chemical potential $\mu_1 \neq \mu_2$. So, there exists a chemical potential difference $\Delta \mu =\mu_2 -\mu_1$ between reservoirs. In this case, it has been shown that the element of density matrix can be found as \cite{78}
\begin{eqnarray}
\rho_{11}&=&(1-L_+)(1-L_-)-\mathcal{R}, \\
\rho_{22}&=& L_-(1-L_+)+\mathcal{R}, \nonumber \\
\rho_{33}&=& L_+(1-L-)+\mathcal{R}, \nonumber \\
\rho_{44}&=& L+L_-\mathcal{R}, \nonumber \\
\rho_{23} &=& -\frac{M_+ + M_-}{2 + i \frac{\Omega}{J}},\nonumber 
\end{eqnarray}
where 
\begin{equation}
\mathcal{R}=\frac{(M_+ + M_-)^2}{4+ (\frac{\Omega}{J})^2}.
\end{equation}

\begin{figure}[!h]
    \centering
  \includegraphics[width = 0.85\linewidth]{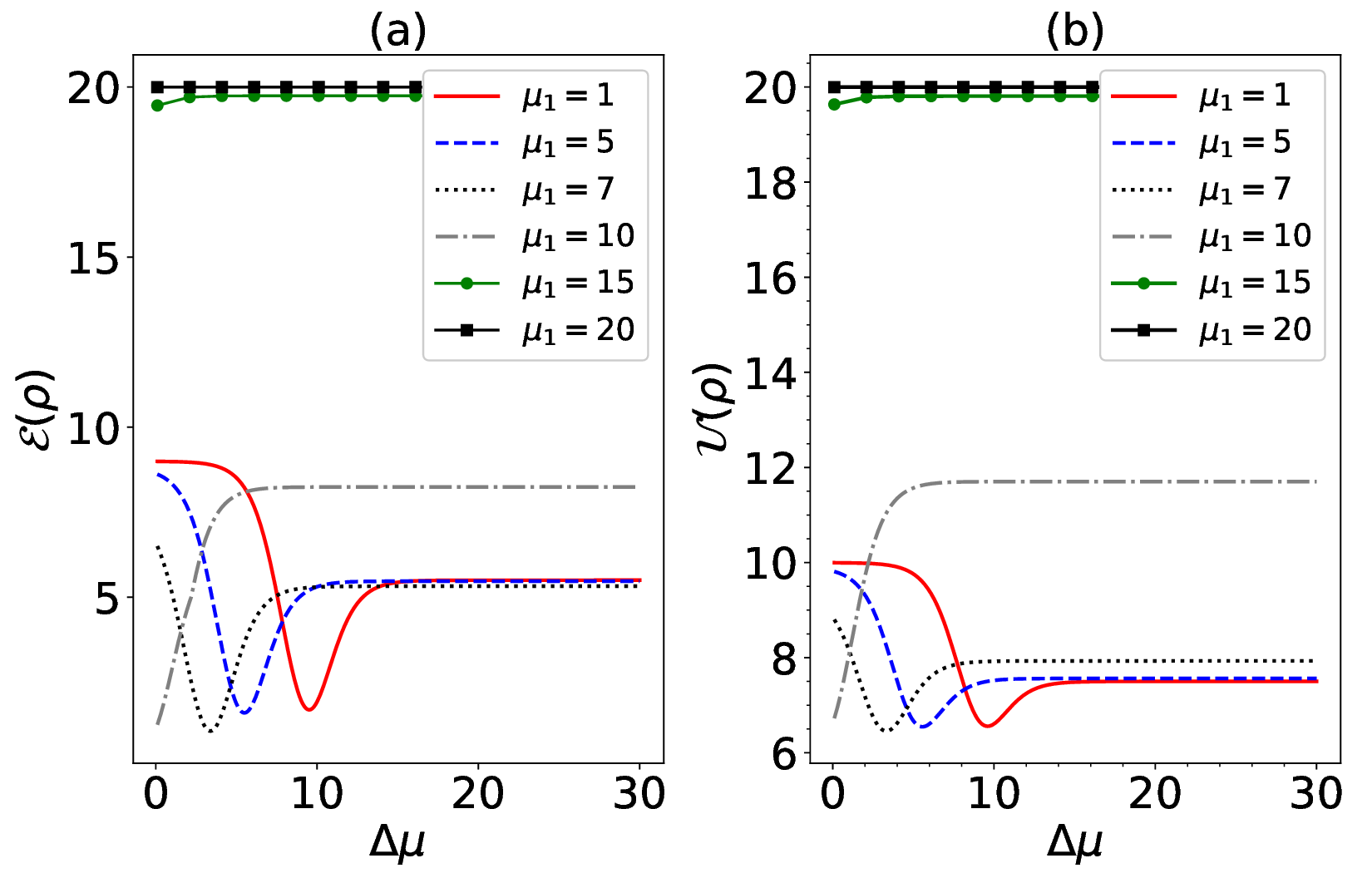}
    \centering
    \caption{(a) The steady-state ergotropy in non-equilibrium fermion reservoirs as a function of $\Delta \mu$ for different value of the base chemical potential $\mu_1$. (b) The internal energy in non-equilibrium boson reservoir as a function of  $\Delta \mu$ for different value of the base chemical potential $\mu_1$. The values of other parameters for both plot are $\omega_1=\omega_2=10$, $J_1=J_2=1$ and $T_1=T_2=1$.}\label{Fig6}
\end{figure}

In Fig.\ref{Fig6}, the steady state ergotropy and internal energy have been plotted as a function of chemical potential difference $\Delta \mu$ for different values of the base chemical potential $\mu_1$. From Fig.\ref{Fig6}(a), it is observed that the steady state ergotropy in non-equilibrium fermion reservoirs has the elaborated behavior in terms of the $\Delta \mu$. This complicated manner depends on the value of the base chemical potential $\mu_1$. For small values of $\mu_1$ ($\mu_1 < \omega$), the ergotropy shows the non-monotonic behavior with $\Delta \mu$, initially decreasing and then increasing as $\Delta \mu$ rises. When $\mu_1$ is larger ($\mu_1 > \omega$), the ergotropy  increases monotonically. If $\mu_1$ becomes sufficiently large, the ergotropy reaches to its maximum constant value.  Overall, the findings demonstrate that increasing the chemical potential beyond the transition frequency of the qubits enhances the efficiency of work extraction from the coupled qubits inside individual reservoirs with different chemical potential. Similar results regarding steady state internal energy, as with ergotropy, can be observed in Fig.\ref{Fig6}(b).
\section{EXTRACTING WORK FROM  ASYMMETRIC QUBITS}\label{sym}
So far, we have studied the maximum extractable work from the coupled symmetric qubits with $\omega_1=\omega_2$ at both bosonic and fermionic reservoirs, with equilibrium and non-equilibrium scenario. Now, we intend to investigate the maximum extractable work from the coupled asymmetric qubits $\omega_1 \neq \omega_2$ at both fermionic and bosonic reservoirs. To understand more details about how to obtain the elements of the steady-state density matrix, see Ref. \cite{78}. The elements of the density matrix are introduced using the following formulation
\begin{equation}
\mathcal{L}_{\pm}=L_\pm \pm M_\pm \cos \theta,\quad \mathcal{M}_\pm = M_\pm \sin \theta.
\end{equation}
 The elements of the density matrix in the antisymmetric situation will be obtained by substituting $\mathcal{L}_\pm$ and $\mathcal{M}_\pm$ instead of $L_\pm$ and $M_\pm$ in the symmetric case. So, the elements of the density matrix for asymmetric qubits coupled to boson reservoir are given by
\begin{eqnarray}
\rho_{11}&=& \frac{1}{\mathcal{G}}\left[ (1+\mathcal{L}_+)(1+\mathcal{L}_-)- s_1 s_2 R \right], \\ 
\rho_{22}&=&  \frac{1}{\mathcal{G}}\left[ \mathcal{L}_-(1+\mathcal{L}_+)+s_2 z_1 R\right], \nonumber \\
\rho_{33}&=&  \frac{1}{\mathcal{G}}\left[ \mathcal{L}_+(1+\mathcal{L}_-)+s_1 z_2 R\right], \nonumber \\
\rho_{44}&=&  \frac{1}{\mathcal{G}}\left[ \mathcal{L}_+\mathcal{L}_--z_1z_2 R \right], \nonumber \\
\rho_{23}&=& \frac{1}{\mathcal{G}} \left[  \frac{\mathcal{M}_+(1+2\mathcal{L}_-)+\mathcal{M}_-(1+2\mathcal{L}_+)}{2(1+\mathcal{L}_++\mathcal{L}_-)+i\frac{\Omega}{J}}\right], \nonumber
\end{eqnarray}

with

\begin{eqnarray}
s_1 &=& \mathcal{M}_+-\mathcal{M}_-(3+2\mathcal{L}_++2 \mathcal{L}_-), \\ 
s_2 &=& \mathcal{M}_--\mathcal{M}_+(3+2\mathcal{L}_++2 \mathcal{L}_-), \nonumber \\
z_1 &=& \mathcal{M}_+ + \mathcal{M}_-(1+2 \mathcal{L}_++2\mathcal{L}_-), \nonumber \\
z_2 &=& \mathcal{M}_- + \mathcal{M}_+(1+2 \mathcal{L}_++2\mathcal{L}_-), \nonumber \\
R &=& \frac{1}{4(1+\mathcal{L}_++\mathcal{L}_-)^2 + (\frac{\Omega}{J})^2}. \nonumber
\end{eqnarray}

where $\mathcal{G}$ is the normalization factor and is given by 

\begin{equation}
\mathcal{G}=(1+2 \mathcal{L}_+)(1+ 2\mathcal{L}_-) 
- 16 \mathcal{M}_+ \mathcal{M}_-(1+\mathcal{L}_++\mathcal{L}_-)^2 R. 
\end{equation}

Similarly, the elements of the density matrix for an asymmetric qubit coupled to the fermion reservoir will be obtained as follows
\begin{eqnarray}
\rho_{11}&=&(1-\mathcal{L}_+)(1-\mathcal{L}_-)-\mathcal{R}, \\
\rho_{22}&=& \mathcal{L}_-(1-\mathcal{L}_+)+\mathcal{R}, \nonumber \\
\rho_{33}&=& \mathcal{L}_+(1-L-)+\mathcal{R}, \nonumber \\
\rho_{44}&=& L+\mathcal{L}_-\mathcal{R}, \nonumber \\
\rho_{23} &=& -\frac{\mathcal{M}_+ + \mathcal{M}_-}{2 + i \frac{\Omega}{J}},\nonumber 
\end{eqnarray}
where 
\begin{equation}
\mathcal{R}=\frac{(\mathcal{M}_+ + \mathcal{M}_-)^2}{4+ (\frac{\Omega}{J})^2}.
\end{equation}

\begin{figure}[!h]
    \centering
  \includegraphics[width = 0.85\linewidth]{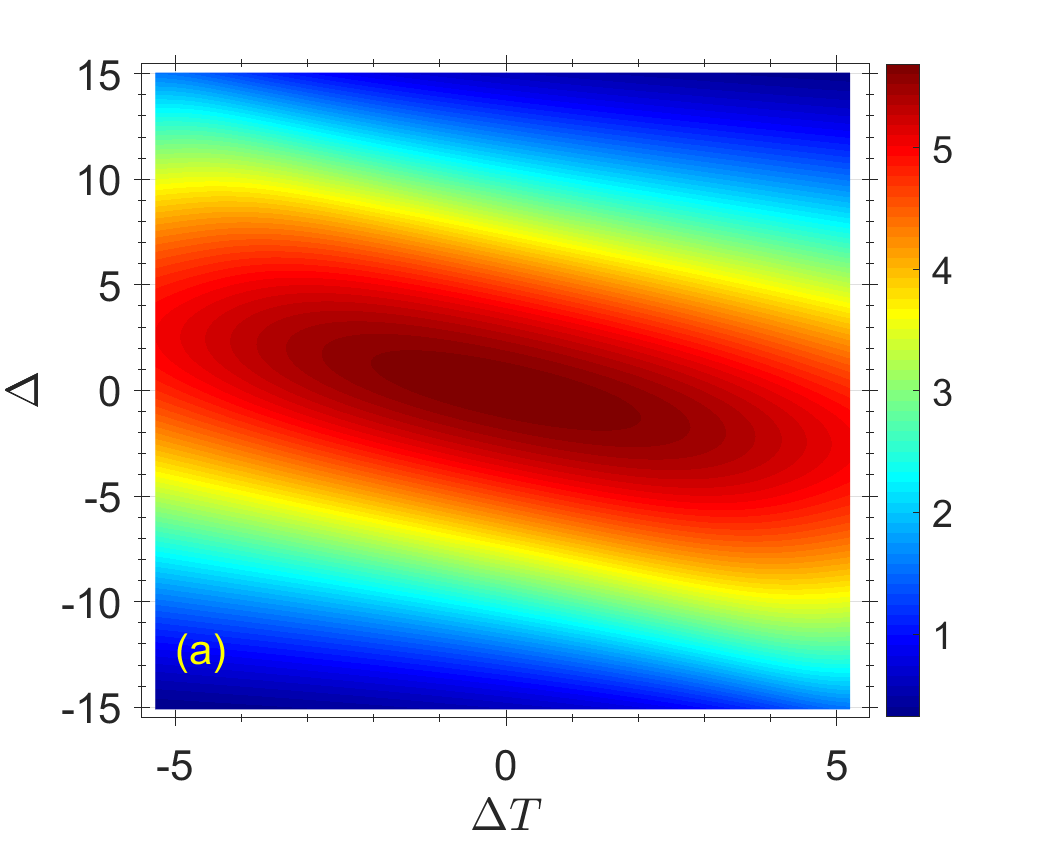}
  \includegraphics[width = 0.85\linewidth]{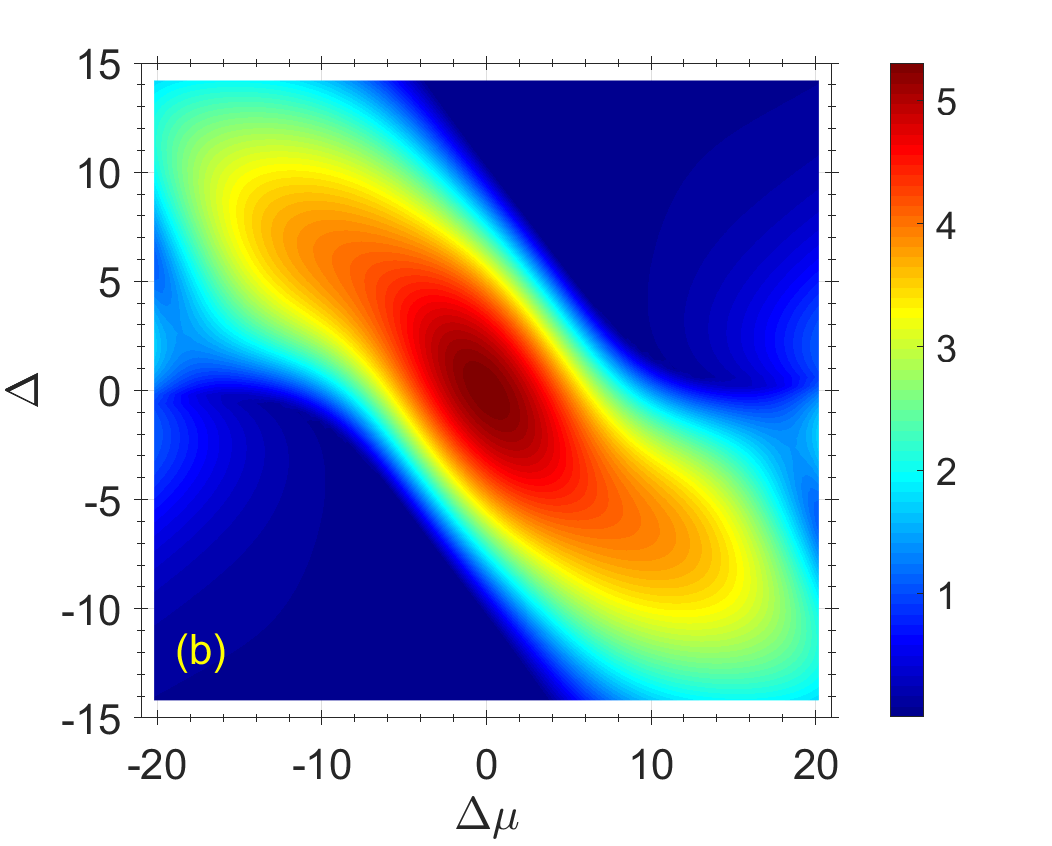}
    \centering
    \caption{(a) The steady-state ergotropy of coupled qubits embedded in boson reservoir  as a function of temperature difference $\Delta T$ and detuning $\Delta$ with $\bar{T}=\frac{T_1+T_2}{2}=3$. (b) The steady-state ergotropy of coupled qubits embedded in fermion reservoir  as a function of chemical potential difference $\Delta \mu$ and detuning $\Delta$ with $\bar{\mu}=\frac{\mu_1+\mu_2}{2}=4$, $T_1=T_2=1.5$. The othe parameters are set as $J_1=J_2=1$, $\lambda=6$ and $\bar{\omega}=\frac{\omega_1+\omega_2}{2}=10$ for both panels.}\label{Fig7}
\end{figure}

Our results about the maximum work extraction of the asymmetric coupled qubits inside boson and fermion reservoirs have been shown in Fig. \ref{Fig7}(a) and Fig. \ref{Fig7}(b), respectively. Fig.\ref{Fig7}, is the counter plot of the steady state maximum work extraction, decribed by ergotropy, as a function of the non-equilibrium parameter $\Delta T$ and $\Delta \mu$ for boson and fermion reservoirs, respectively and the  detuning between transition frequency of the coupled qubits  $\Delta=\omega_1 - \omega_2$ of the coupled qubits. 

As shown in Fig.\ref{Fig7}(a) , ergotropy is at its maximum value in the center of the phase diagram, i.e., under conditions where the boson reservoirs are in equilibrium situation and the qubits are symmetric.  Therefore, it can be concluded that the maximum work will be extracted under conditions where the coupled qubits are symmetric and situated in the equilibrium reservoirs. It is also observed that the ergotropy has its minimum values at the
top right and bottom left corners of the phase diagram. In the
top right corner we have $\Delta >0$ and $\Delta T >0$ which are associated with $\omega_1 > \omega_2$ and $T_1<T_2$, respectively, while in the bottom left corner $\omega_1<\omega_2$ and $T_1 > T_2$. So, it can be concluded that putting the qubit with a higher transition frequency in a bosonic environment at a lower temperature does not lead to an optimal work extraction. From Fig.\ref{Fig7}(a), it can be seen the the ergotropy has its maximum value in the center of the phase diagram, i.e., under conditions where the fermion reservoirs are in equilibrium situation and the qubits are symmetric.  Therefore, it can be concluded that the maximum work will be extracted under conditions where the coupled qubits are symmetric and situated in the equilibrium fermion reservoirs. It is also observed that the ergotropy has its minimum values at the top right and bottom left corners of the phase diagram. In the
top right corner we have $\Delta >0$ and $\Delta \mu >0$ which are associated with $\omega_1 > \omega_2$ and $\mu_1< \mu_2$, respectively, while in the bottom left corner  $\omega_1<\omega_2$ and $\mu_1 > \mu_2$. So, it can be concluded that putting the qubit with a higher transition frequency in a fermion reservoir at a lower chemical potential does not lead to an optimal work extraction.
\section{SUMMARY AND RESULTS}\label{conclusion}
In this work, we studied the steady state work extraction from the two coupled qubits interacting with its individual reservoirs. We demonstrated that when the qubits are placed in equilibrium boson reservoirs, the inter-qubits interaction strength and the temperature of the reservoirs have a detrimental effect on the maximum extractable work  from the qubits. The situation is different for the case in which the coupled qubits embedded in equilibrium boson reservoirs. In this case the maximum extractable work grows monotonically when the chemical potential greater than transition frequency of the qubits. When coupled qubits placed in non-equilibrium boson reservoirs the ergotropy  is suppressed by increasing the temperature difference between reservoirs. The behavior of the maximum extractable work from coupled qubit in non-equilibrium fermion reservoirs is complicated. In this case the behavior of the ergotropy depends on the base chemical potential. The ergotropy has non-monotonic behavior when the base chemical potential smaller the the transition frequency of the qubits. In such a way that ergotropy decreases and then grows monotonically with the reservoirs chemical potential difference. However, when the base chemical potential greater than transition frequency of the qubits the ergotropy grows monotonically with the reservoirs chemical potential difference and reaches to constant value for large value of the reservoirs chemical potential difference. In this work, the impact of symmetric or asymmetric qubits on the maximum extractable work from the qubits was also studied. It was observed that under conditions where the coupled qubits are symmetric and placed in equilibrium boson or fermion reservoirs, the work extraction occurs more optimally. It is also observed that putting the qubit with a higher transition frequency in a bosonic environment at a lower temperature does not lead to an optimal work extraction. Similarly, the work extraction is not optimal when the qubit with a higher transition frequency embedded in a fermion reservoir at a lower chemical potential.


\end{document}